\documentclass[12pt]{article}
\usepackage{amsmath,amssymb}
\usepackage[dvips]{graphicx}
\usepackage{braket}

\newcommand{\de}{\partial}
\newcommand{\be}{\begin{equation}}
\newcommand{\ba}{\begin{eqnarray}}
\newcommand{\ea}{\end{eqnarray}}
\newcommand{\ee}{\end{equation}}

\newcommand{\f}{\frac}
\newcommand{\s}{\sqrt}

\newcommand{\ddd}{\cdot\cdot\cdot}
\newcommand{\no}{\nonumber \\}

\begin{document}

\begin{titlepage}
\thispagestyle{empty}

\begin{flushright}
YITP-13-115
\\
IPMU13-0215
\\
\end{flushright}


\begin{center}
\noindent{\large \textbf{Volume Law for the Entanglement Entropy in Non-local QFTs}}\\
\vspace{2cm}

Noburo Shiba $^{a}$
and Tadashi Takayanagi $^{a,b}$
\vspace{1cm}

{\it
 $^{a}$Yukawa Institute for Theoretical Physics (YITP),\\
Kyoto University, Kyoto 606-8502, Japan\\
$^{b}$Kavli Institute for the Physics and Mathematics of the Universe,\\
University of Tokyo, Kashiwa, Chiba 277-8582, Japan\\
}

\vskip 2em
\end{center}

\begin{abstract}
In this paper, we present a simple class of non-local field theories whose ground state entanglement entropy follows a volume law as long as the size of subsystem is smaller than a certain scale. We will confirm this volume law both from numerical calculations and from analytical estimation. This behavior fits nicely with holographic results for spacetimes whose curvatures are much smaller than AdS spaces such as those in the flat spacetime.
\end{abstract}

\end{titlepage}

\newpage

\section{Introduction}

The entanglement entropy offers us a universal measure of the degrees of freedom in any quantum many-body systems.  Since it is defined by focusing on an arbitrary subsystem of a given quantum system, we can probe effective degrees of freedom for any fixed length scale and position. Thus this includes quite a lot of information about any ground state.

It has been well-established that in any local quantum field theory with a ultraviolet (UV) fixed point, the entanglement entropy follows the universal rule called area law \cite{Ereview}. This claims that the entanglement entropy $S_\Omega$ for a subsystem $\Omega$ has a UV divergence in the continuum limit of quantum field theories and that the coefficient of this divergence is proportional to the area of the boundary
$\de \Omega$ of the subsystem.

The area law was first found in free field theories \cite{Bombelli:1986rw,Sr}. One way to confirm the area law for interacting field theories is to employ the AdS/CFT correspondence \cite{Maldacena}. The AdS/CFT correspondence argues that a gravitational theory on $d+1$ dimensional anti de Sitter space (AdS space) is equivalent to a $d$ dimensional conformal field theory (CFT), where the latter is typically described by a strongly coupled and large $N$ gauge theory. The holographic formula of entanglement entropy \cite{RT,HRT} shows that the area law holds for such a strongly coupled CFT with a UV fixed point and this heavily relies on the geometry of the AdS space.

The AdS/CFT can be regarded as an example of a more general and earlier idea called holography \cite{Hol}. This principle conjectures that a given gravity theory in a $d+1$ dimensional spacetime ${\cal M}$ is equivalent to a certain quantum many-body system which lives on the $d$ dimensional boundary of $\de{\cal M}$. Therefore it is natural to ask what we can say about general holography from the quantum entanglement viewpoint. If we consider, for example, a flat spacetime in any dimension, we can immediately find that its holographic entanglement entropy satisfies a volume law instead of the area law \cite{LiTa}.
 Refer also to related earlier works \cite{Bar} and recent discussions \cite{Fi,Kar} in gravity duals of non-commutative field theories, where the holographic entanglement entropy was shown to follow a volume law. Therefore it seems important to find a class of field theories whose entanglement entropy satisfies the volume law.

It is well-known that for the generic excited states in any quantum many-body systems, the entanglement entropy satisfies the volume law \cite{Page}. This means that the state which follows the area law is very special. Indeed, the local quantum field theories, which have the area law property, are clearly special in that the interactions are very short range and this property crucially helps to reduce the amount of entanglement.

In lattice models, it is not so difficult to construct models with a volume law. For example, we can consider a spin system with random interactions between
any two pairs of spins. The non-local random interactions obviously lead to a highly entangled ground state which satisfies the volume law. Moreover it has been pointed out that even without local interactions, we can construct lattice models with a volume law if we give up the translational invariance \cite{Lat}.

The purpose of this paper is to present a field theory model with the translational invariance whose ground state satisfies the volume law. The previous arguments suggest that we may get the volume law if we consider suitable non-local field theories.
There have already been suggestions of such field theories in \cite{LiTa,NRT} via heuristic discussions. Also we would like to mention that the paper \cite{longrangeEE} studied milder non-local field theories, which do not lead to the volume law but have modified coefficients of logarithmic divergent term in two dimensional field theories.

In this paper, we will present simple and concrete examples of non-local and non-relativistic free scalar field theories. We will show manifestly that they indeed have the property of volume law both by explicit numerical calculations and by analytical estimations. We will also see that holographic calculations confirm the same behavior.

The paper is organized as follows: In section 2, we will briefly review how to calculate the entanglement entropy in free bosonic quantum many-body systems. In section 3, we will explain our models of non-local scalar field theories in two dimensions and their lattice regularizations. In section 4, we will present numerical results of entanglement entropy for our non-local scalar models in two dimensions
and give their analytical explanations. In section 5, we will generalize our results into higher dimensions. In section 6 we present a holographic interpretation and find that it is consistent with our field theoretic results. In section 7 we summarize our conclusion.

When we were writing the draft of this paper, we noted the paper \cite{Ka}, where the authors showed that the volume law can be obtained for a non-commutative field theory on a fuzzy sphere, which is also an example of non-local field theory (refer also to \cite{Dou} for earlier works on entanglement entropy in non-commutative field theories).

\section{How to compute entanglement entropy: Real time approach} \label {review}
In this section we review the method of computing the entanglement entropy in free field theories developed by Bombelli et al \cite{Bombelli:1986rw}, which we will employ in this paper. Refer to \cite{Sr,Pes,CH,CFH,ANT,CHR,HeSp,HeNi,ShibaBH} for examples of other useful computational methods for the entanglement entropy in free field theories.
As a model amenable to unambiguous calculation we deal with the scalar field
as a collection of coupled oscillators on a lattice of space points,
labeled by capital Latin indices, the displacement at each point giving the value of the scalar field there.
In this case the Hamiltonian can be given by
\begin{equation}
H=\dfrac{1}{2}\delta_{MN}P_M P_N + \dfrac{1}{2} V_{MN} q_M q_N , \label{eq:1-1}
\end{equation}
where $q_M$ gives the displacement of the $M$-th oscillator
and $P_M$ is the conjugate momentum to $q_M$.
The matrix $V_{MN}$ is symmetric and positive definite.
The matrix $V_{MN}$ is independent of $q^M$ and $\dot q^M$.
We can obtain the ground state wave function as
\begin{equation}
\psi (\{q_A\})=\left( \det \dfrac{W}{\pi}\right)^{1/4} \exp (-\dfrac{1}{2}  W_{AB} q_A q_B ), \label{eq:1-2}
\end{equation}
where 
\begin{equation}
W \equiv V^{1/2}.  \label{eq:1-3}
\end{equation}
The matrix $W_{MN}$ is symmetric and positive definite.

Now consider a subsystem (or subregion) $\Omega$ in the space. 
The oscillators in this region will be specified by  lowercase Latin letters,
and those in its complement $\Omega ^c$ will be specified by Greek letters.
We will use the following notation
\begin{alignat}{2}
W_{AB} = \begin{pmatrix}
             W_{ab} & W_{a\beta}  \\
             W_{\alpha b} & W_{\alpha \beta}
             \end{pmatrix}
             \equiv   \begin{pmatrix}
             A & B  \\
             B^T & C
             \end{pmatrix}  &   ~~~~~~
 W^{-1}_{AB} = \begin{pmatrix}
             W^{-1}_{ab} & W^{-1}_{a\beta}  \\
             W^{-1}_{\alpha b} & W^{-1}_{\alpha \beta} \end{pmatrix}
              \equiv   \begin{pmatrix}
             D & E  \\
             E^T & F
             \end{pmatrix}
 \label{eq:1-4}
\end{alignat}
From $W W^{-1}=1$, we have
\begin{alignat}{2}
 \begin{pmatrix}
             1 & 0  \\
             0 & 1
             \end{pmatrix}
             =   \begin{pmatrix}
             A & B  \\
             B^T & C
             \end{pmatrix}
   \begin{pmatrix}
             D & E  \\
             E^T & F
             \end{pmatrix}
             =     \begin{pmatrix}
            A D+BE^T & AE+BF  \\
             B^T D+CE^T & B^TE+ CF
             \end{pmatrix}
. \label{eq:1-5}
\end{alignat}

We can obtain a reduced density matrix
$\rho_{red}$ for $\Omega$ by integrating out over $q^{\alpha }\in \mathbb{R}$ for each of the oscillators in $\Omega^c$, and then we have
\begin{equation}
\rho_{red} (\{q^1_a\} , \{q^2_{b}\}) =
\left( \dfrac{\det \dfrac{W}{\pi} }{\det \dfrac{C}{\pi}} \right)^{1/2}
\exp [ -\dfrac{1}{2} (q^1_a,q^2_b)
\begin{pmatrix}
             X & 2Y  \\
             2Y & X
             \end{pmatrix}
\begin{pmatrix}
             q^1_a  \\
             q^2_b
             \end{pmatrix} ] \label{eq:1-6}
\end{equation}
where $X=A-\tfrac{1}{2} B C^{-1} B^{T}, Y=-\tfrac{1}{4}B C^{-1} B^{T}$.

We can write the density matrix as one for non coupled degrees of freedom
by making an appropriate linear transformation on $q_a$.
Finally the entanglement entropy $S_{\Omega}=-tr \rho_{red} \ln \rho_{red} $ is given by \cite{Bombelli:1986rw}
\begin{gather}
 S_{\Omega}=     \sum_n f(\lambda_n) ,  \label{eq:new3-1}  \\
   f(\lambda) \equiv   \ln (\dfrac{1}{2} \lambda ^{1/2} ) + (1+\lambda  )^{1/2 } \ln [(1+\lambda  ^{-1})^{1/2 } + \lambda ^{-1/2} ]   ,
     \label{eq:1-7}
\end{gather}
where $\lambda_n$ are the eigenvalues of the matrix
\begin{equation}
\Lambda ^a _{~b} = - W^{a\alpha} W_{\alpha b} =-(E B^T)^a_{~~b} =(DA)^a_{~~b} -\delta^a_{~~b} .    \label{eq:1-8}
\end{equation}
In the last equality we have used (\ref{eq:1-5}).
The last expression in (\ref{eq:1-8}) is useful for numerical calculations when $\Omega$ is smaller than $\Omega^c$, 
because the indices of $A$ and $D$ take over only the space points on $\Omega$ and the matrix sizes of $A$ and $D$ are smaller than those of $B$ and $E$ as emphasized in \cite{Shiba}.
It can be shown that all of $\lambda_n$ are non-negative as follows.
From (\ref{eq:1-5}) we have
\begin{equation}
A \Lambda = -AEB^T = BFB^T .    \label{eq:1-9}
\end{equation}
It is easy to show that $A,C,D$ and $F$ are positive definite matrices when $W$ and $W^{-1}$ are positive definite matrices.
Then $A\Lambda$ is a positive semidefinite matrix as can be seen from (\ref{eq:1-9}).
So all eigenvalues of $\Lambda$ are non-negative.
After all, we can obtain the entanglement entropy by solving the eigenvalue problem of $\Lambda $.

\section{Two dimensional non-local scalar fields on lattices} \label {lattice}
We apply the above formalism to free scalar fields in $(1+1)$-dimensional Minkowski spacetime.
As an ultraviolet regulator, we replace the continuous space coordinate $x$ by a lattice of discrete points with spacing $a$.
As an infrared cutoff, we allow $n \equiv x/a$ to take only a finite integer values $-N/2<n \leq N/2$.
Outside this range we assume the lattice is periodic.
Later we will take $N$ to infinity.
The dimensionless Hamiltonian $H_0\equiv a H$ is given by
\begin{equation}
H_0\equiv a H 
 \equiv \sum_{n} \dfrac{1}{2} \pi_{n}^2 + \sum_{m,n} \dfrac{1}{2} \phi_{m} V_{mn} \phi_{n} , \label{eq:2-1}
\end{equation}
where $\phi_{n}$ and $\pi_{n}$ are dimensionless and Hermitian, and obey the canonical commutation relations
\begin{equation}
[\phi_{n} , \pi_{m}]=i \delta_{nm}  . \label{eq:2-2}
\end{equation}

As an example, let us consider the Klein Gordon field whose mass is $m$.
We can diagonalize the matrix $V$ by a Fourier transform \cite{creutz1985quarks} and obtain
\begin{equation}
(V_{KG})_{m n}= N^{-1} \sum_{k} [a^2 m^2+2  (1-\cos \dfrac{ 2\pi k }{N} ) ] e^{2\pi i k(n-m)/N} , \label{eq:2-3}
\end{equation}
\begin{equation}
(W_{KG})_{m n}= N^{-1} \sum_{k} [a^2 m^2+2 (1-\cos \dfrac{ 2\pi k }{N} ) ]^{1/2} e^{2\pi i k(n-m)/N} , \label{eq:2-4}
\end{equation}
\begin{equation}
(W_{KG})_{m n}^{-1}= N^{-1} \sum_{k} [a^2 m^2+2 (1-\cos \dfrac{ 2\pi k }{N} ) ]^{-1/2} e^{2\pi i k(n-m)/N} , \label{eq:2-5}
\end{equation}
where the index $k$ is also an integer in the range of $-N/2<k \leq N/2 $.
We take $N$ to infinity and change the momentum sum into an integral with the replacements $q=2\pi k /N$ and
$N^{-1} \sum_{k}\rightarrow \int_{-\pi}^{\pi} \tfrac{d q}{(2\pi )}$,
and then we have
\begin{equation}
(W_{KG})_{m n}= \int_{-\pi}^{\pi} \dfrac{d q}{(2\pi )} e^{i q (n-m)} [a^2 m^2 + 2 (1-\cos q ) ]^{\tfrac{1}{2}}  , \label{eq:2-6}
\end{equation}
\begin{equation}
(W_{KG})_{m n}^{-1}= \int_{-\pi}^{\pi} \dfrac{d q}{(2\pi )^d} e^{i q (n-m)} [a^2 m^2 + 2  (1-\cos q ) ]^{\tfrac{-1}{2}}  . \label{eq:2-7}
\end{equation}
Then the laplacian on lattices is $ 2 (1-\cos q )$.

Next we turn to non-local scalar fields theories which we are interested in this paper.
The Hamiltonian is defined by
\begin{equation}
H=\f{1}{2}\int dx   \left[ (d\phi/dt)^2 + B_0 \phi e^{A_0 (-\partial^2)^{w/2} } \phi \right],
\label{eq:2-8}
\end{equation}
where $A_0,B_0$ are positive constants.
We define dimensionless constants
\be
B_0=B/a^2,\ \ \ A_0=a^w A.
\ee
We can change $B$ into $1$ by rescaling $t$.
Thus the entanglement entropy is independent of $B$ and we can set $B=1$.
We obtain $W$ and $W^{-1}$ as follows
\begin{equation}
(W^w)_{mn}=\int_{-\pi}^{\pi} \dfrac{dq}{2\pi} e^{iq(n-m)}
 \exp [A/2(2-2\cos q )^{w/2} ]
\label{eq:2-9}
\end{equation}
\begin{equation}
(W^w)^{-1}_{mn}=\int_{-\pi}^{\pi} \dfrac{dq}{2\pi} e^{iq(n-m)}
 \exp [-A/2(2-2\cos q )^{w/2} ]
\label{eq:2-10}
\end{equation}
For later convenience we define
\begin{equation}
W_n=W_{m,m+n}, \ \ \ W^{-1}_n=W^{-1}_{m,m+n}.
\label{eq:2-11}
\end{equation}
When $w=1,2$, we can calculate $W,W^{-1}$ analytically as we will show below.

\subsection{Case1: $w=1$}
In the case $w=1$, we obtain
\begin{equation}
\begin{split}
(W^{w=1})_{n}&=\int_{-\pi}^{\pi} \dfrac{dq}{2\pi} e^{iqn}
 \exp [A/2(2-2\cos q )^{1/2} ]
=\int_{-\pi}^{\pi} \dfrac{dq}{2\pi} e^{iqn}
 \exp [A |\sin (q/2)| ] \\
&=\int_{0}^{2\pi} \dfrac{dq}{2\pi} e^{iqn}
 \exp [A \sin (q/2) ]
=\int_{0}^{\pi} \dfrac{dx}{\pi} e^{2ixn}
 \exp [A \sin x ] \\
&=\int_{0}^{\pi} \dfrac{dx}{\pi}[\cos(2nx-iA\sin x) +i\sin(2nx-iA\sin x)] \\
&=J_{2n}(iA) +i \mathbf{E}_{2n}(iA)
\end{split}
\label{eq:2-12}
\end{equation}
where $J_{2n}$ is the Bessel function and $\mathbf{E}_{2n}$ is the Weber function.
We can rewrite (\ref{eq:2-12}) as
\begin{equation}
(W^{w=1})_{n}=(-1)^n I_{2n}(A) +\dfrac{1}{2}A(-1)^n  _1\tilde{F}_2 [1;(3-2n)/2,(2n+3)/2;A^2/4]
\label{eq:2-13}
\end{equation}
where $I_{2n}$ is the modified Bessel function and we have used
$J_{2n}(iA)=i^{2n} I_{2n} (A)=(-1)^n  I_{2n} (A)$ and
\begin{equation}
\mathbf{E}_{2n}(iA)=-\dfrac{1}{2}iA(-1)^n  _1\tilde{F}_2 [1;(3-2n)/2,(2n+3)/2;A^2/4].
\label{eq:2-14}
\end{equation}
Here $_1\tilde{F}_2$ is the regularized hypergeometric function.
The  regularized hypergeometric function is defined as
\begin{equation}
  _p\tilde{F}_q [a_1, \dots , a_p ; b_1, \dots , b_q ; z] \equiv \dfrac{_p F_q [a_1, \dots , a_p ; b_1, \dots , b_q ; z]   }{ \Gamma (b_1) \dots \Gamma (b_q) } ,
\label{eq:2-14-2}
\end{equation}
where $_p F_q [a_1, \dots , a_p ; b_1, \dots , b_q ; z]  $ is the  hypergeometric function.
This expression is manifestly real and suitable for numerical calculations.
We can obtain $W^{-1}$ by replacing $A\rightarrow -A$ in $W$:
\begin{equation}
(W^{w=1})^{-1}_{n}=(-1)^n I_{2n}(A)  -\dfrac{1}{2}A(-1)^n  _1\tilde{F}_2 [1;(3-2n)/2,(2n+3)/2;A^2/4].
\label{eq:2-15}
\end{equation}

\subsection{Case2: $w=2$}
In the case $w=2$ we find
\begin{equation}
\begin{split}
(W^{w=2})_{n}&=\int_{-\pi}^{\pi} \dfrac{dq}{2\pi} e^{iqn}
 \exp [A(1-\cos q ) ]
=e^A \int_{-\pi}^{\pi} \dfrac{dq}{2\pi} e^{iqn}
 \exp [-A \cos q ] \\
&=e^A \int_{0}^{\pi} \dfrac{dq}{\pi} \cos(nq)
 \exp [-A \cos q ]
=e^A i^n J_n(iA)=e^A (-1)^n I_n(A).
\end{split}
\label{eq:2-16}
\end{equation}
We can obtain $W^{-1}$ by replacing $A\rightarrow -A$ in $W$:
\begin{equation}
(W^{w=2})^{-1}_{n}=e^{-A} (-1)^n I_n(-A)=e^{-A} I_n(A).
\label{eq:2-17}
\end{equation}

\section{Computations of Entanglement Entropy}

Now we would like to turn to the main part of this paper: computations of entanglement entropy for our non-local scalar field theories. We will first present numerical results which support the volume law and later give an analytical explanation.

\subsection{Numerical calculations}
We perform matrix operations and calculate the eigenvalues $\lambda_n$ of the matrix $\Lambda $ in (\ref{eq:1-8}). Then finally we can obtain the entanglement entropy in  (\ref{eq:1-7}) with \texttt{Mathematica 8}. We define the subsystem $\Omega$ to be an interval on the one dimensional lattice with the length $L$. Since the columns and rows of the matrix $\Lambda $ describe points in $\Omega$, the size of the matrix $\Lambda$ is given by $L\times L$.

\begin{figure}
 \includegraphics[width=8cm,clip]{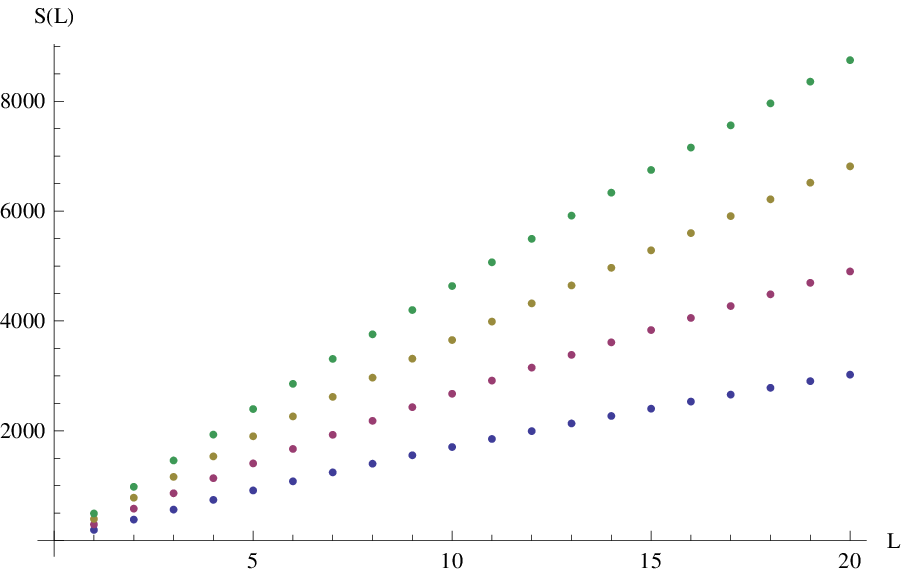}
 \hspace{5mm}
 \includegraphics[width=8cm,clip]{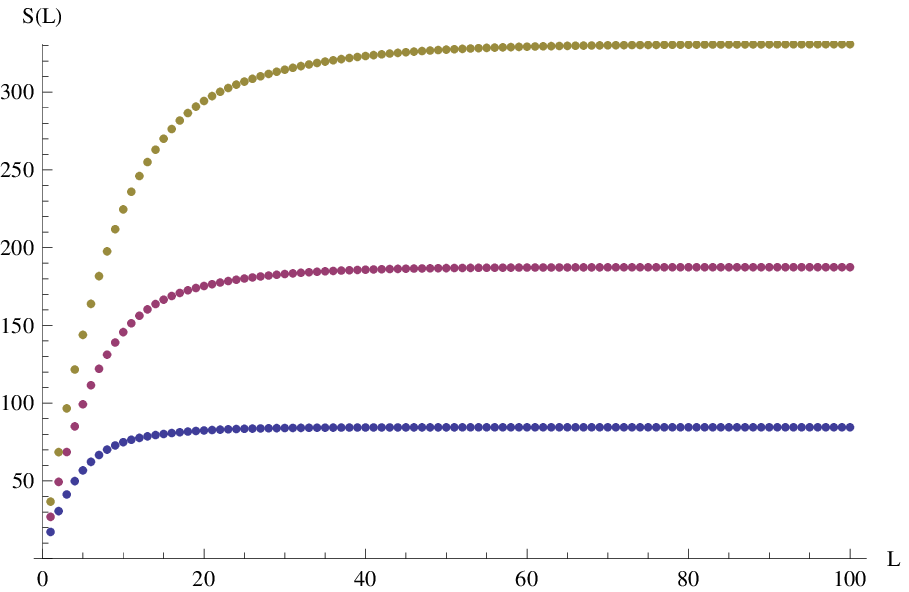}%
 \caption{The entanglement entropy $S_{\Omega}(L)$ of one interval whose length is $L$ for $w=1$
as a function of $L$. In the left picture, the blue, red, yellow and green points correspond to $A=400, 600, 800, 1000$. In the right picture, the blue, red and yellow points correspond to $A=40, 60, 80$.}
 \label{S(L)w1l}
 \end{figure}

For $w=1$ (Case 1), we show the computed values of $S_{\Omega}(L)$
as a function of $L$ in Fig.\ref{S(L)w1l}.
As can be seen, $S_{\Omega}(L)$ is proportional to $L$ when $L<<A$ and approaches its maximum value when $L>>A$.
By using the data between $1\leq L\leq 20$ for $2000 \leq A \leq3000$, we obtain $S_{\Omega}(L)\simeq 0.48AL$
for $L<<A$.
By using the data for $100\leq A \leq200$, we obtain $S_{\Omega}(L)\simeq 0.055 A^{2.0}$
for $L>>A$.

\begin{figure}
 \includegraphics[width=8cm,clip]{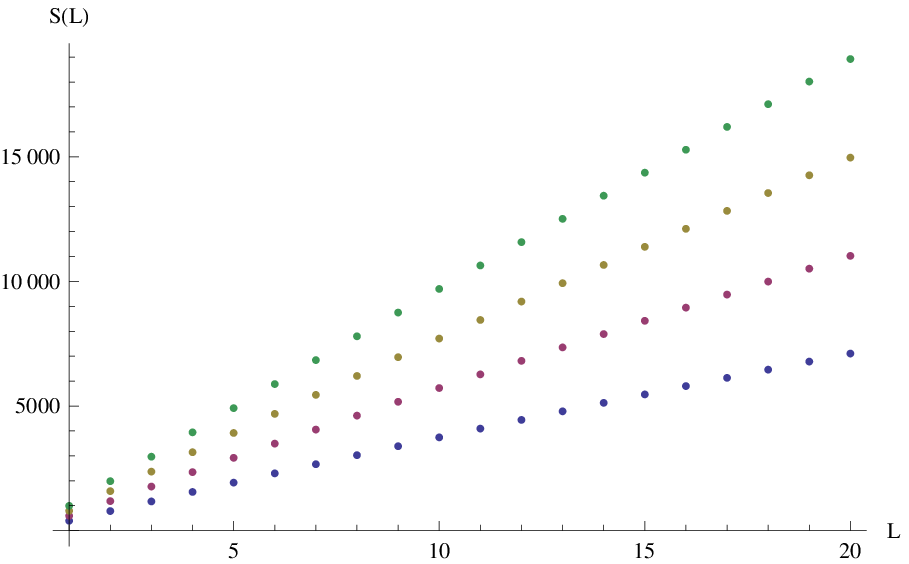}%
 \hspace{5mm}
 \includegraphics[width=8cm,clip]{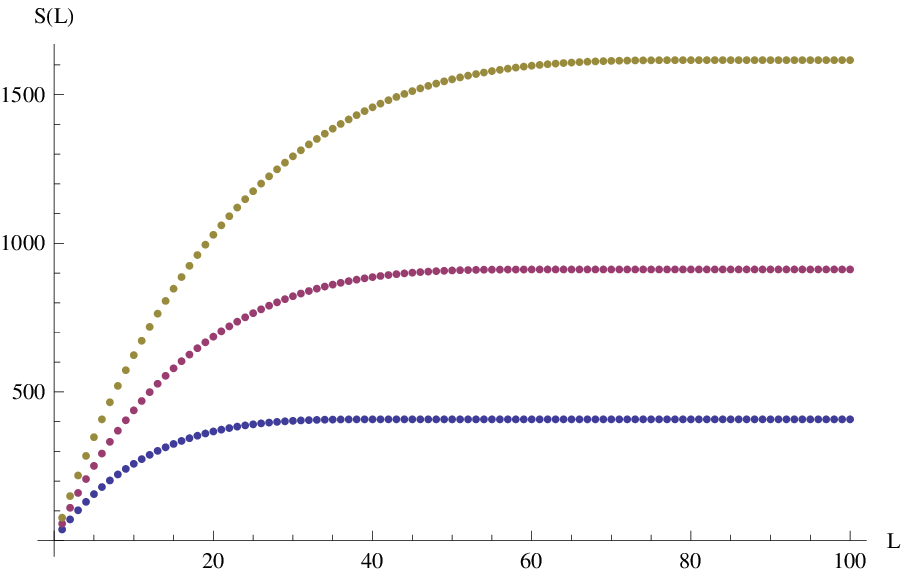}%
 \caption{The entanglement entropy $S_{\Omega}(L)$ of one interval whose length is $L$ for $w=2$
as a function of $L$. In the left picture, the blue, red, yellow and green points correspond to $A=400, 600, 800, 1000$. In the right picture, the blue, red and yellow points correspond to $A=40, 60, 80$.}
 \label{S(L)w2l}
 \end{figure}

For $w=2$ (Case 2), we show the computed values of $S_{\Omega}(L)$
as a function of $L$ in Fig.\ref{S(L)w2l}.
As can be seen, $S_{\Omega}(L)$ is proportional to $L$ when $L<<A$ and approaches its maximum value when $L>>A$. 
This behavior is similar to the entanglement entropy for $w=1$.
By using the data between  $1\leq L\leq 20$ for $2000 \leq A \leq3000$,
we obtain $S_{\Omega}(L)\simeq 0.98AL$
for $L<<A$.
By using the data for $100\leq A \leq200$, we obtain $S_{\Omega}(L)\simeq 0.26 A^{2.0}$
for $L>>A$.

These numerical result suggests that the entanglement entropy behaves like
\ba
S_\Omega(L)&\simeq& c_1 LA\ \ (L<<A),\no
&\simeq& c_2 A^2\ \ (L>>A),
\label{finr}
\ea
where $c_1$ and $c_2$ are order one constants which depends only on the value of $w$.
Our numerical results implies the identification $c_1=w/2$. In this way, we can conclude that ground states of these models satisfy the volume law as long as the subsystem size is small as $L<<A$.

\subsection{Analytical Explanation}


We consider the behavior of the entanglement entropy by examining the matrix $\Lambda$  in (\ref{eq:1-8}).
We can write explicitly $\Lambda$ as
\begin{equation}
\Lambda_{m,n} =\sum_{l=1}^{L} W^{-1}_{m-l} W_{l-n} - \delta_{m,n}
\label{eq:3-1}
\end{equation}
where $1\leq m,n \leq L$.
First we consider the region $L\ll A$.
In this region we can use the asymptotic expansion of $W_n$ and $W^{-1}_n$
for large values of $A$.
From (\ref{eq:2-12}), (\ref{eq:2-15}), (\ref{eq:2-16}) and (\ref{eq:2-17}), we have the asymptotic expansions for $A\gg 1$ as
\begin{equation}
(W^{w=1})_{n} \sim (-1)^n e^A \sqrt{ \dfrac{2}{\pi A} } [1-\dfrac{1}{A} (2 n^2 -\dfrac{1}{8} )+\dots ] \label{eq:3-2}
\end{equation}

\begin{equation}
(W^{w=1})_{n}^{-1} \sim \dfrac{2}{\pi A} [1-\dfrac{1}{A^2} (4 n^2 -1)+\dots]  \label{eq:3-3}
\end{equation}

\begin{equation}
(W^{w=2})_{n} \sim (-1)^n \dfrac{e^{2A}}{\sqrt{2\pi A}} [1-\dfrac{1}{2A}(n^2 -\dfrac{1}{4} ) +\dots ]  \label{eq:3-4}
\end{equation}

\begin{equation}
(W^{w=2})_{n}^{-1} \sim \dfrac{1}{\sqrt{2\pi A}} [1-\dfrac{1}{2A}(n^2 -\dfrac{1}{4} ) +\dots ]
\label{eq:3-5}
\end{equation}

From these asymptotic expansions, $\Lambda_{m,n} \sim e^{wA}$ and the magnitude of
nonzero eigenvalues $\lambda_i$ is $e^{wA}$.
From (\ref{eq:1-7}) each $\lambda_i$ contribute $\f{wA}{2}$ to the entanglement
entropy. Since all of $L$ eigenvalues are expected to contribute, we obtain $S_{\Omega}(L) \simeq \f{w}{2} LA=c_1LA$. 

\begin{figure}
 \includegraphics[width=8cm,clip]{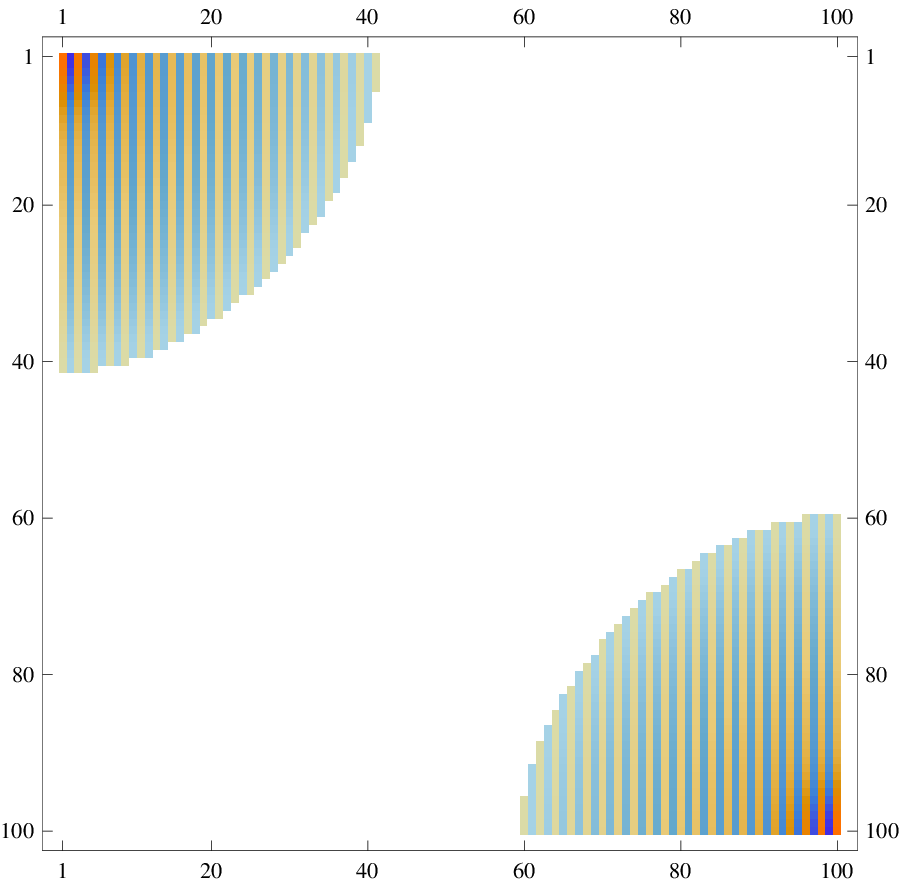}%
 \hspace{5mm}
 \includegraphics[width=8cm,clip]{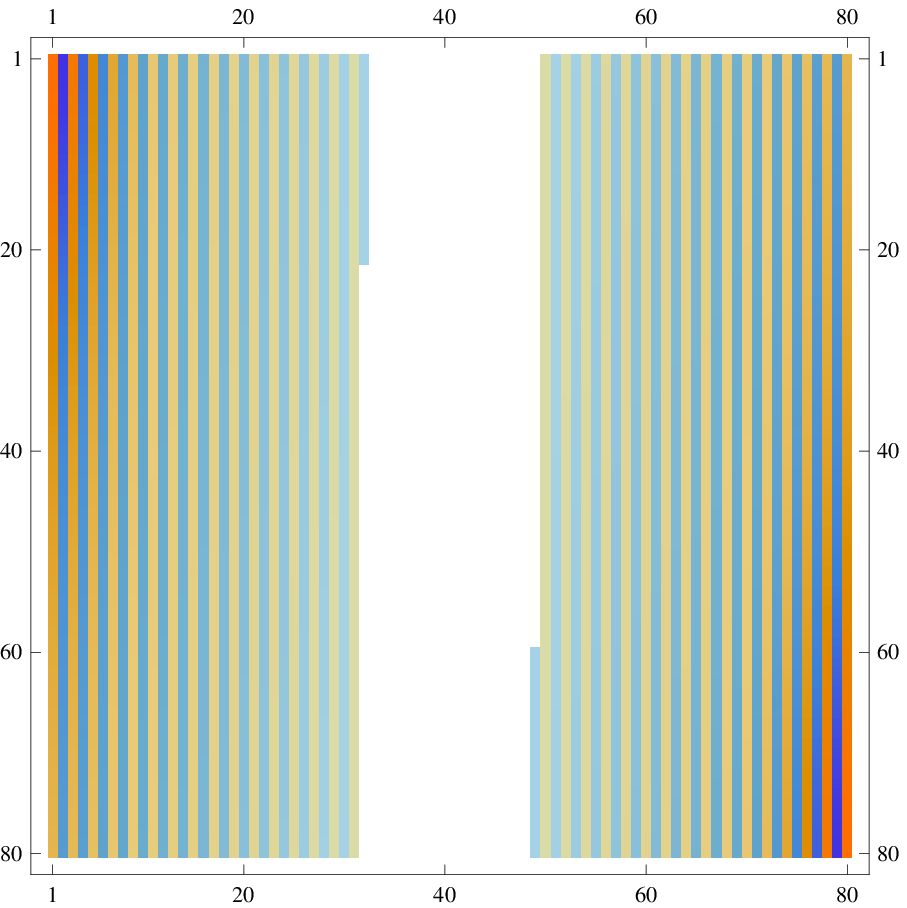}%
 \caption{The matrices $\Lambda(L,A)$ for $w=2$ (left) and $w=1$ (right).
In the left picture
we show $\Lambda(L=100,A=20)$ for $w=2$ and the magnitude of the matrix elements in the white region is smaller than $10^{-15}$ times the maximum of the matrix elements.
The matrix elements in the orange (blue) region is positive (negative).
In the right picture
we show $\Lambda(L=80,A=50)$ for $w=1$ and the magnitude of the matrix elements in the white region is smaller than $10^{-18}$ times the maximum of the matrix elements.
The matrix elements in the orange (blue) region are positive (negative) again. }
 \label{matlambda}
 \end{figure}

Next we consider the region $L\gg A$.
In this case we use the asymptotic forms of $W_n$ and $W_n^{-1}$ for large $n$.
First we consider the case $w=2$.
We can obtain the asymptotic form of $I_n(A)$ in (\ref{eq:2-16}) and (\ref{eq:2-17}) from the integral representation

\begin{equation}
I_n (A)=\dfrac{1}{2\pi i} \left( \int_{-\infty-\pi i}^{-\pi i}
+\int_{-\pi i}^{\pi i} +\int_{\pi i}^{\pi i + \infty} \right)
e^{A \cosh t -nt}dt.
\label{eq:3-6}
\end{equation}
By using the method of steepest descent, we obtain
\begin{equation}
I_n (A) \sim \sqrt{ \dfrac{ \tanh \beta }{2 \pi} } e^{-n(\beta -\coth \beta )}
\label{eq:3-7}
\end{equation}
where $\sinh \beta = n/A  $ and $n \gg A$.
From the asymptotic form of $I_{n}(A)$,
we can see that $|W_n|$ and $|W_n^{-1}|$ decrease rapidly when $n(\gg A)$ increases.
From  (\ref{eq:2-16}), (\ref{eq:2-17}) and (\ref{eq:3-1})
we obtain  $\Lambda_{m,n} \simeq 0$ when $|m-n| \gg A$.
Furthermore we can see that $\Lambda_{m,n} \simeq 0$ when
$A \ll m,n \ll L-A$ by the following argument.
By using the identity $\sum_{l=-\infty}^{\infty} W_{m-l}^{-1} W_{l-n}=\delta_{m,n}$,
we can rewrite $\Lambda$ in (\ref{eq:3-1}) as
\begin{equation}
\Lambda_{m,n} =-\left( \sum_{l=- \infty}^{0} +\sum_{l=L+1}^{\infty} \right) W^{-1}_{m-l} W_{l-n}  .
\label{rew lambda}
\end{equation}
Notice $W_{l-n}$ in the sum in the above expression.
The largest $W_{l-n}$ in the sum is $W_{min ( n, L+1-n )}$ and
$ min ( n, L+1-n ) \gg A$ when $A \ll n \ll L-A$.
So $\Lambda_{m,n} \simeq 0$ when $A \ll n \ll L-A$.
In the same way we can see that $\Lambda_{m,n} \simeq 0$ when $A \ll m \ll L-A$.
Finally, $\Lambda_{m,n}$ are significant values only when $m,n \lesssim A$ or $L-A \lesssim m,n$.
We show $\Lambda$ in Fig \ref{matlambda}.
The blocks in which $\Lambda_{m,n}$ are significant values are $A \times A$ matrices and
$\Lambda_{m,n} \sim e^{2A} $ in these blocks.
In the same way as in the case $L \ll A$, we can estimate the entanglement entropy and
obtain $S_{\Omega}(L) \simeq c_2 A^2$ when $L \gg A$.

In the case $w=1$, we can estimate the entanglement entropy in the similar way.
In this case we can see numericaly that $W_n$ decrease faster than $W_n^{-1}$.
Via the same argument under (\ref{rew lambda}) we can see that
 $\Lambda_{m,n} \simeq 0$ when $A \ll n \ll L-A$.
We show $\Lambda$ in Fig\ref{matlambda}.
We can rewrite the entanglement entropy in (\ref{eq:new3-1}) as $S_{\Omega}=\mbox{Tr} f(\Lambda)$.
From the form of the matrix $\Lambda$, we can see that Tr$\Lambda^l$ is independent of $L$.
Thus the entanglement entropy is constant when $L\gg A$.

\section{Higher Dimensional Generalization}

Next we would like to consider a straightforward generalization of our two dimensional scalar field model (\ref{eq:2-8}) to the $d$ dimensional one ($d>2$) defined by the Hamiltonian
\be
H=\int d^{d-1}x\left[\f{1}{2}(\de_t\phi)^2+B_0 \phi\cdot e^{A_0(-\de_i\de_i)^{w/2}}\cdot\phi\right], \label{Hamg}
\ee
where we defined $\de_i=\f{\de}{\de x_i}$ and $x_i\ (i=1,2,\ddd,d-1)$ are the coordinates of ${\mathbb{R}}^{d-1}$.

We divide the coordinates of ${\mathbb{R}}^{d-1}$ into two parts: $x_1\in {\mathbb{R}}$
and $(x_2,\ddd,x_{d-1})\in {\mathbb{R}}^{d-2}$. We take the Fourier transformation with respect to the latter and obtain
\ba
&& H=\int d^{d-2}k\int dx_1  \left[\f{1}{2}\de_t\phi(k)\de_t\phi(-k)+B_0 \phi(k)\cdot e^{A_0(-\de^2+k^2)^{w/2}}\cdot\phi(-k)\right], \no
&& \equiv \int d^{d-2}k~ H(k).
\ea
In this way we can decompose the Hamiltonian as a sum of two dimensional scalar Hamiltonian $H(k)$ over the transverse momenta $k$. Note that in the Hamiltonian $H(k)$,
$|k|$ plays the role of mass parameter in the two dimensional theory as is familiar in the  Kaluza-Klein theories. As a IR regularization, we compactify the space ${\mathbb{R}}^{d-2}$ into a torus with the radius $Ra$ ($a$ was the lattice constant and $R$ is the size of torus in the lattice space.). Then the momentum is quantized as $k_i=2\pi\f{n_i}{Ra}$, where $n_i$ runs $-\f{R}{2}\leq n_i< \f{R}{2}$. We choose the subsystem $\Omega$ in the definition of entanglement entropy $S_\Omega$ to be a strip with the width $La$ defined by
\be
-\f{La}{2}\leq x_1\leq \f{La}{2},\ \ \ 0\leq x_2,\ddd,x_{d-1}\leq Ra. \label{stripd}
\ee

First consider the case $w=2$. The entanglement entropy $S_{\Omega}$ for the ground state of the Hamiltonian $H(k)$ is identical to our original model in two dimension i.e. $H(0)$. This is because the $k$ dependence only appears as a factor $e^{A_0k^2}$ in front of $\phi^2$ term, which does not change our calculation of $S_{\Omega}$ as is clear from our previous calculations. Therefore we can estimate the total contribution when $L<<A$ as follows
\be
S_{\Omega}\simeq \prod_{i=2}^{d-1}\sum_{n_i=-R/2}^{R/2} AL=A L R^{d-2}.
\ee
Therefore we confirmed the volume law in any dimension.

Now let us move on to more general $w$. We define $p$ to be the momentum in the $x_1$ direction. Since $1<<L<<R$ and $p\sim \f{1}{La}<<\f{1}{a}$, the dominant contribution comes from the region
$k>>p$. When $k>>p$, we find
\be
e^{A_0(p^2+k^2)^{w/2}}\simeq e^{A_0k^w}e^{\f{A_0 w}{2}k^{w-2}p^2}.
\ee
Thus we can approximate the calculation of $S_{\Omega}$ by that of $w=2$ with $A_0$ replaced with $\f{A_0 w}{2}k^{w-2}$ for each $k$. In this way we can estimate as follows
\ba
S_{\Omega}\simeq \prod_{i=2}^{d-1}\sum_{n_i=-R/2}^{R/2} \f{A_0 w}{2a^2}\left(\f{2\pi |n|}{Ra}\right)^{w-2} L\simeq C_1\cdot ALR^{d-2},
\ea
where $C_1$ is a certain order one constant; we employed the same definition $A=A_0 a^{-w}$ of $A$ as in the two dimensional case. In this way, again we found the volume law in any dimension.

We can similarly analyze $S_{\Omega}$ in our $d$ dimensional theory even when $L>>A$.
In the end we find $S_{\Omega}\simeq C_2 A^2 R^{d-2}$, for a certain constant $C_2$. In summary, for our $d$ dimensional non-local scalar field model, we obtained the following behavior:
\ba
S_\Omega(L)&\simeq& C_1 ALR^{d-2}\ \ (L<<A),\no
&\simeq& C_2 A^2R^{d-2}\ \ (L>>A).
\label{finrr}
\ea

In this way we confirmed that our higher dimensional model also has the property of volume law for a small size subsystem. On the other hand, when the size of $\Omega$ gets larger than the parameter $A$, it follows an area law.

\section{Holographic Interpretation}

Finally we would like to discuss a holographic counterpart of our field theory analysis. Originally, the non-local scalar field model defined by the Hamiltonian  (\ref{Hamg}) was considered in \cite{NRT} in the context of an interpretation of AdS/CFT correspondence
as an entanglement renormalization. See also \cite{LiTa} for a similar but different model which was proposed as a toy model of holographic dual of gravity in a flat spacetime.

In \cite{Swingle}, it has been conjectured that
 a framework of real space renormalization, called MERA (multi-scale entanglement renormalization ansatz) \cite{MERAR}, is equivalent to the AdS/CFT correspondence.
This idea allows us to relate the entanglement structure of a quantum state in MERA to the metric of its gravity dual.

Especially, by using the continuum limit of MERA (called cMERA \cite{cMERA}), a formula for the metric in the extra dimension has been proposed in \cite{NRT}. Consider the metric in the $d+1$ dimensional gravity dual:
\be
ds^2\propto g_{uu}du^2+\f{e^{2u}}{a^2}\sum_{i=1}^{d-1}dx_i^2+g_{tt}dt^2, \label{metmera}
\ee
where $a$ is the UV cut off (or lattice spacing) as in our previous sections and $u$ is the coordinate of extra direction. We regard $u=0$ as the boundary of $d+1$ dimensional spacetime where its holographic dual lives. Note that we ignored any constant factor of the metric which depends only on the Hamiltonian of the theory.

If we consider the free scalar field theory defined by (\ref{Hamg}), then the proposed formula \cite{NRT} for the dual metric from the viewpoint of cMERA
computes $g_{uu}$ as follows
\be
g_{uu}=\f{A_0^2}{a^2}e^{2wu}. \label{guu}
\ee

Now we introduce the standard extra dimension coordinate $z\equiv ae^{-u}$ and then we can rewrite the metric (\ref{metmera}) as
\be
ds^2\propto A_0^2\f{dz^2}{z^{2(w+1)}}+\f{1}{z^2}\sum_{i=1}^{d-1}dx_i^2+g_{tt}dt^2.
\ee
Moreover, it is useful to define the coordinate $y=z^{-w}$ to rewrite the spacial part of the above metric into
 \be
 ds^2_{space}\propto A_0^2 dy^2 +y^{\f{2}{w}}\sum_{i=1}^{d-1}dx_i^2, \label{smets}
 \ee
 where rescaled $x_i$ by a finite amount.

Now we would like to study the holographic entanglement entropy \cite{RT} for
the gravity dual (\ref{smets}). We choose the subsystem $\Omega$ to be the strip defined by (\ref{stripd}), with the understanding that $R$ is infinitely large by taking the decompactifying limit.  The holographic entanglement entropy is given by
\be
S^{hol}_\Omega=\f{1}{4G_N}\mbox{Area}(\gamma_\Omega),
\ee
where the $d-1$ dimensional surface $\gamma_\Omega$ is the minimal surface which ends on $\de \Omega$ (i.e.
$\de \gamma_\Omega=\de \Omega$) \cite{RT}; $G_N$ is the Newton constant of the $d+1$ dimensional gravity.
The area of minimal surface ending on $\de \Omega$ can be obtained by minimizing
\be
\mbox{Area}=R^{d-2}a^{d-2}\int^{La/2}_{-La/2}dx_1 y^{\f{d-2}{w}}\s{A_0^2 y'^2+y^{\f{2}{w}}},
\ee
where $y'=\f{\de y}{\de x_1}$. By deriving a conserved quantity (`Hamiltonian'), we find
\be
A_0\f{dy}{dx_1}=y^{1/w}\s{y^{2(d-1)/w}/y^{2(d-1)/w}_*-1}, \label{wer}
\ee
where $y_*$ is the integration constant.
We assumed that the minimal surface extends between $y_*\leq y<y_{\infty}$, where
$y_{\infty}=a^{-w}$ is the counterpart of the UV cut off in the gravity dual, which corresponds to the boundary of $d+1$ dimensional spacetime defined by $u=0$. Also $y=y_*$ is the turning point of the surface where $y'$ vanishes.
By integrating (\ref{wer}) we find
\be
a^{w-1}\cdot\int^{y_{\infty}}_{y_*} \f{dy}{y^{1/w}\s{y^{2(d-1)/w}/y^{2(d-1)/w}_*-1}}=\f{L}{2A}.
\label{rati}
\ee
In the end, the minimal area is expressed as the integral
\be
\mbox{Area}=2R^{d-2}a^{d-2}A_0\int^{y_{\infty}}_{y_*} dy
\f{y^{(2d-3)/w}y^{-(d-1)/w}_*}{\s{y^{2(d-1)/w}/y^{2(d-1)/w}_*-1}}.  \label{harea}
\ee

First we are focusing on the region $L<<A$ as we assumed to show the volume law in our previous sections. From (\ref{rati}), this case corresponds to
$y_*\simeq y_{\infty}(=a^{-w})$ and thus the minimal surface is always very close to the boundary $y=y_{\infty}$. Thus the minimal area (\ref{harea}) is proportional to the volume $LR^{d-2}$ of the $d-1$ dimensional space where the non-local scalar field lives. Indeed, in this case, we can estimate the holographic entanglement entropy as follows
\be
S_{\Omega}\propto R^{d-2}a^{d-1}L y_{\infty}^{(d-1)/w}=LR^{d-2},
\ee
where we omitted the coefficient which only depends on the theory (or equally Hamiltonian).

It is also intriguing to ask the behavior of holographic entanglement entropy when
$L>>A$. First note that there exist two disconnected minimal surfaces which are simply given by $x_1=\pm La/2$ with $0\leq y\leq y_{\infty}$. The sum of these two disconnected surfaces is another candidate of minimal surfaces $\gamma_{\Omega}$ for the holographic entanglement entropy $S_\Omega$. Indeed since the metric in the $x_i$ direction vanishes at $y=0$, it satisfies the required condition $\de \gamma_{\Omega}=\de \Omega$. The minimal area principle of holographic entanglement entropy tells us that we should choose $\gamma_{\Omega}$ to be the sum of the disconnected surfaces when $L>>A$. This leads to the estimate
 Area$(\gamma_{\Omega})\propto AR^{d-2}$. Notice also that in the opposite case $L<<A$,
 we have to choose $\gamma_{\Omega}$ to be the connected surface because the area of the connected one is clearly smaller than that of the disconnected ones.
 This is a typical example of `phase transition' for the entanglement entropy as observed in a variety of holographic examples (see e.g. \cite{NiTa,KKM,He}), which is considered to be an artifact of large $N$ limit.

In this way, our holographic results confirmed the behavior (\ref{finr}) and (\ref{finrr}), assuming that the proportionality coefficient, which we are not able to fix, is given by $A$ times a numerical constant. Especially this supports the claim that the entanglement entropy satisfies the volume law when $L<<A$.

\section{Conclusion}

In this paper we presented a simple class of non-relativistic field theories whose entanglement entropy satisfies a volume law as long as the size of subsystem is smaller than a certain parameter (called $A$) of the theory, which parameterizes the magnitude of non-locality.
These field theories are highly non-local in real space and this is obviously the reason why it follows the volume law rather than the area law. This model has another parameter $w$ which is related to the types of non-locality we are considering.

We confirmed our model follows the volume law when $w=1$ and $w=2$ in the two dimensional scalar field theory both from numerical calculations and analytical estimates. We also extended this result into higher dimensions. The final result of entanglement entropy $S_\Omega$, when the subsystem $\Omega$ is a strip with the width $L$, is summarized in (\ref{finrr}). Also our holographic calculation agrees with these field theory results and furthermore predicts that we will obtain the volume law for any values of $w(>0)$.

It will be intriguing a future problem to go beyond free field theories by taking into interactions as well as to extend our constructions to fermions. Another interesting direction is to pursuit holography for general spacetimes by using entanglement entropy.
It is an important future problem to better understand our holographic relation
between almost flat spacetimes and non-local field theories.

\section*{Acknowledgements}

We would like to thank Masahiro Nozaki, Shinsei Ryu and Erik Tonni for useful discussions.
NS and TT are supported by JSPS Grant-in-Aid for Scientific
Research (B) No.25287058 and JSPS Grant-in-Aid for Challenging
Exploratory Research No.24654057. TT is also
supported by World Premier International
Research Center Initiative (WPI Initiative) from the Japan Ministry
of Education, Culture, Sports, Science and Technology (MEXT).

\end{document}